\documentclass[twocolumn]{aastex631}

\usepackage{savesym}
\savesymbol{tablenum}
\usepackage{siunitx}
\DeclareSIUnit\Msun{\ensuremath{\mathrm{M}_{\odot}}}
\DeclareSIUnit\erg{\ensuremath{\mathrm{erg}}}
\DeclareSIUnit\angstrom{\text {Å}}
\restoresymbol{SIX}{tablenum}

\usepackage[version=3]{mhchem}

\graphicspath{{./}{figures/}}

\hypersetup{linkcolor=red,citecolor=magenta,urlcolor=blue}

\newcommand{\ngc}{NGC\,6397}
\newcommand{\tuc}{47\ Tucanae}
\newcommand{\omegacen}{$\omega$\,Centauri}
\newcommand{\teff}{T_\mathrm{eff}}
\newcommand{\logg}{\log_{10}(g)}
\newcommand{\review}[1]{#1}

\begin{document}

\title{JWST Imaging of the Closest Globular Clusters\,--\,II. Discovery of Brown Dwarfs in \ngc{} and Measurement of Age from the Brown Dwarf Cooling Sequence, using SANDee -- a New Grid of Model Isochrones across the Hydrogen-Burning Limit}

\author[0000-0003-0398-639X]{Roman Gerasimov}
\affiliation{Department of Physics and Astronomy, University of Notre Dame, Nieuwland Science Hall, Notre Dame, Indiana 46556, USA}

\author[0000-0003-4080-6466]{Luigi R.\ Bedin}
\affiliation{Istituto Nazionale di Astrofisica, Osservatorio Astronomico di Padova, Vicolo dell’Osservatorio 5, I-35122, Padova, Italy}

\author[0000-0002-6523-9536]{Adam J.\ Burgasser}
\affiliation{Department of Astronomy and Astrophysics, University of California, San Diego, La Jolla, California 92093, USA}

\author[0000-0003-3714-5855]{Daniel Apai}
\affiliation{Department of Astronomy and Steward Observatory, The University of Arizona, 933 N. Cherry Avenue, Tucson, Arizona 85721, USA}
\affiliation{Lunar and Planetary Laboratory, The University of Arizona, 1629 E. University Blvd., Tucson, Arizona 85721, USA}

\author[0000-0003-1149-3659]{Domenico Nardiello}
\affiliation{Dipartimento di Fisica e Astronomia, Universit{\`a} di Padova, Vicolo dell’Osservatorio 3, I-35122, Padova, Italy}
\affiliation{Istituto Nazionale di Astrofisica, Osservatorio Astronomico di Padova, Vicolo dell’Osservatorio 5, I-35122, Padova, Italy}

\author[0009-0005-8159-8490]{Efrain Alvarado III}
\affiliation{Department of Astronomy, University of California, Berkeley, California 94720-3411, USA}

\author{Jay Anderson}
\affiliation{Space Telescope Science Institute, 3700 San Martin Drive, Baltimore, Maryland 21218, USA
}

\begin{abstract}

Globular clusters contain vast repositories of metal-poor stars that represent some of the oldest stellar generations in the Universe. The archaeological footprint of early Galactic evolution may be retained in the measurable properties of globular clusters, such as their ages, mass functions and chemical abundances. Until recently, all photometric studies of globular clusters were restricted to stellar members. Now, the sensitivity of JWST can extend this analysis to the substellar regime. If detected in sufficient numbers, brown dwarf members can provide tight constraints on the properties of their parent population. We present \texttt{SANDee} -- a new grid of stellar models that accurately represent the color-magnitude diagrams of globular clusters across the hydrogen-burning limit at a wide range of metallicities. Using JWST NIRCam photometry and the new models, we identify three brown dwarfs in the globular cluster \ngc{} with $\teff=1300-\qty{1800}{\K}$, confirmed by both proper motion and model fitting. We use the observed luminosities of discovered brown dwarfs to obtain the first age estimate of a globular cluster from its substellar cooling sequence: $13.4\pm\qty{3.3}{Gyr}$. We also derive the local mass function of the cluster across the hydrogen-burning limit and find it to be top-heavy, suggesting extensive dynamical evolution. We expect that the constraints on both age and mass function of \ngc{} derived in this work can be greatly improved by a second epoch of NIRCam imaging in the same field.

\end{abstract}


\section{Introduction} \label{sec:introduction}
Brown dwarfs are objects with masses intermediate between those of stars and planets. Unlike planets, brown dwarfs are produced by the star formation process; unlike stars, they are insufficiently massive to sustain steady-state hydrogen burning in their cores. The boundary between stars and brown dwarfs, the hydrogen-burning limit, is $\sim\qty{0.07}{\Msun}=75\ \mathrm{M}_\mathrm{jup}$ for solar-like chemistries \citep{HBL_1,HBL_2}, but may be \review{over $25\%$} higher at lower metallicities \citep{HBL_3,HBL_4}. The boundary between brown dwarfs and planets should be determined by the formation process rather than initial mass, but since the origins of an object may be challenging to establish, the boundary is often approximated as the deuterium-burning limit at $\sim13\ \mathrm{M}_\mathrm{jup}$ \citep{DBL_1,DBL_2}. Brown dwarfs account for over $20\%$ of the local stellar census \citep{davy_20pc}, and this estimate is likely to increase in the future, as the lower mass limit of star/brown dwarf formation has not been identified even in the deepest, most complete volume-limited surveys \citep{BD_abundance,BD_abundance_2,davy_20pc}.

Unlike main sequence stars, brown dwarfs do not have a usable mass-temperature relationship, as they do not attain energy equilibrium and, instead, gradually evolve along cooling curves determined by the initial mass and chemistry \citep{cooling_curves}. The resulting mass-age degeneracy makes the measurement of brown dwarf masses particularly challenging. It is thus not surprising that the earliest unambiguously confirmed brown dwarfs \citep{early_BD_1,early_BD_2} had effective temperatures far below those attainable even by the lowest-mass hydrogen burning stars ($\teff\lesssim\qty{1200}{\K}$). However, young brown dwarfs that have not had sufficient time to cool may have temperatures comparable to or even exceeding those of main sequence stars, further complicating their identification \citep{M45_hot_BD,TW_Hya_hot_BD}. The maximum age that would qualify as ``young'' in this context is determined by the proximity of the mass of the brown dwarf to the hydrogen-burning limit, and may in principle be arbitrarily old for an object very close to the limit \citep{BD_close_to_HBL}. Therefore, the distributions of temperatures and luminosities in a population of brown dwarfs are also strongly dependent on the initial mass function that determines the expected number of objects per mass interval \citep{slow_BDs_1,slow_BDs_2,adam_gap}.

While the sensitivity of substellar populations to the mass function and age may pose a challenge to the characterization of isolated brown dwarfs in the field, it also provides a means to diagnose mass functions and ages in coeval stellar populations through brown dwarf cooling sequences. The differentiation between steady-state stellar and cooling substellar members is especially pronounced in older populations, providing extended baselines in stellar parameters. Globular clusters are the oldest ($\gtrsim \qty{12}{Gyr}$, \citealt{GC_ages,GC_ages_2}) and largest ($\gtrsim 10^5$ members, \citealt{GC_masses}) coeval stellar populations in the Galaxy, and are therefore prime targets for this analysis. Nonetheless, at the large distances ($\gtrsim\qty{5}{kpc}$, \citealt{GC_distances}) to globular clusters, the faint luminosities and cool temperatures of brown dwarfs have, until recently, restricted observational studies to stellar members.

The James Webb Space Telescope (JWST) is the first facility capable of revealing the brown dwarf cooling sequences in nearby globular clusters. In this study, we present the discovery of brown dwarfs in the globular cluster \ngc{} (Lac III.11, \citealt{lacaille_catalog}), and use their observed properties to infer the age and mass function of the parent population. Situated at $\qty{2.5}{kpc}$ \citep{NGC6397_distance}, \ngc{} is the second closest globular cluster after M4; yet, due to the low metallicity ($[\ce{Fe/H}]\sim-2$, \citealt{matteo_NGC6397}) of \ngc{}, the apparent brightness of its brown dwarf cooling sequence \review{may exceed} that of M4 or any other globular cluster\footnote{The new stellar models presented in Section~\ref{sec:isochrones} of this paper suggest that the hydrogen-burning limit in \ngc{} is brighter \review{than that in M4 by $0.34\qty{-0.37}{mag}$ and $0.02-\qty{0.04}{mag}$ in the \texttt{F150W2} and \texttt{F322W2} bands of JWST NIRCam, respectively, assuming M4 is \qty{0.6}{kpc} closer and $0.9-\qty{1.1}{dex}$ more metal-rich} than \ngc{}, all other parameters being equal (extinction at infrared wavelengths is much smaller than the effect of metallicity).}.

Initial and present-day mass functions of stellar populations provide the most direct insight into star formation and subsequent dynamical evolution. The mass functions of globular clusters are of special interest, as the origin of these objects is not well-established (e.g. see the review in \citealt{GC_formation_review}, the discussion of the globular cluster metallicity floor in \citealt{GC_zscale_floor_1,GC_zscale_floor_2}, and their possible pre-galactic nature, \citealt{pregalactic_GC_1,pregalactic_GC_2}). Proposed theories of chemical enrichment in globular clusters are known to be in conflict with the inferred mass functions (the so-called \textit{mass budget problem}, \citealt{mass_deficit_1,mass_deficit_2,mass_deficit_3}). Furthermore, due to their old ages and low metallicities, globular clusters may serve as probes of the poorly understood transition between the bottom-heavy mass functions of Population~I and II stars \citep{Kroupa,Kroupa_2}, and the predicted top-heavy mass function of Population~III stars \citep{popIII_IMF_1,popIII_IMF_2,popIII_IMF_3}.

While present-day mass functions of globular clusters are widely diverse \citep{sollima_1,NGC6397_MF_2}, it cannot be ruled out that the observed spread is largely driven by dynamical evolution from a universal initial mass function \citep{MF_vs_environment,sollima_1,NGC6397_MF_2,harvey_6397,IMF_from_MF,sollima_2}.
Observations of brown dwarfs in the field suggest the possibility that the substellar initial mass function may be regulated by a distinct mechanism from its higher-mass counterpart \citep{IMF_anomaly_1,IMF_anomaly_2,IMF_anomaly_3}. If so, extending mass function measurements in globular clusters beyond the hydrogen-burning limit may alleviate the degeneracy between the initial mass function and dynamical evolution.

As the oldest coeval populations known, globular clusters provide a direct constraint on the age of the Universe \citep{GC_age_universe_0,GC_ages_0,GC_age_universe_3,GC_age_universe,GC_age_universe_2} and retain the dynamical and chemical footprint of the early evolution of galaxies \citep{GC_archaeology,GC_archaeology_2,GC_archaeology_3,GC_archaeology_4}. In order to fully utilize the potential of globular clusters as ``galactic clocks'', precise estimates of their ages are required. Globular clusters are most commonly dated through isochrone fitting \review{to photometric observations of main sequence turnoff and post-main sequence stars}. This approach suffers from degeneracies between the age of the cluster and other parameters, including interstellar reddening, \review{chemistry}, and the convective properties of stars \citep{GC_age_universe_0,GC_ages_2,GC_age_universe_2}. While random errors in globular cluster ages on the order of $\ll\qty{0.5}{Gyr}$ are readily attainable (e.g., \citealt{precise_GC_ages}), systematic errors may still exceed $\qty{1}{Gyr}$ (e.g., see Table~3 in \citealt{renno_2020}). It is therefore imperative to evaluate the inferred globular cluster ages against alternative dating techniques with independent systematics. For \ngc{}, the white dwarf cooling sequence suggests a relatively young age of $11.5\pm\qty{0.5}{Gyr}$ \citep{ngc6397_wd_age}, isochrone fitting of the main sequence turnoff presents an older age of $12.6\pm\qty{0.7}{Gyr}$ \citep{matteo_NGC6397}, and beryllium dating of individual members yields an even older age of $\gtrsim \qty{13.3}{Gyr}$ \citep{beryllium_dating}.

The continuous cooling of brown dwarf members allows for a new approach to dating globular clusters. At the age of $\sim\qty{0.5}{Gyr}$, globular clusters are expected to develop a scarcely populated \textit{stellar/substellar gap} in their luminosity functions that separates the lowest-mass hydrogen burning stars and the highest-mass brown dwarfs \citep{adam_gap,roman_omega_cen,roman_47Tuc}. The gap proceeds to grow with age, attaining a width of several magnitudes at the typical ages of Milky Way globular clusters. In this study, we demonstrate that the ages of globular clusters can be inferred from the density of observed objects within the stellar/substellar gap, even if the overall number of identified brown dwarfs is very small.

Prior to the launch of JWST, the stellar/substellar gap prevented even the deepest photometric surveys from detecting the brown dwarf cooling sequence. While the end of the main sequence in \ngc{} was unambiguously seen in Hubble Space Telescope (HST) images \citep{ngc_6397_king,harvey_6397}, no members were detected at fainter magnitudes. In the globular cluster M4, dedicated searches for brown dwarf candidates in deep HST fields \citep{BD_hunt_1,BD_hunt_2} identified \review{four brown dwarf candidates beyond the end of the main sequence, but were only able to confirm cluster membership for one of them (\textit{BD2}), and found its photometric properties to be more consistent with the white dwarf sequence}. The first tentative detection of a brown dwarf in a globular cluster was reported by \citet{rolly_BD} using the Near Infrared Camera (NIRCam) on JWST. This candidate substellar member of the globular cluster \tuc{} -- \textit{BD10} -- was found to have a luminosity consistent with the expected onset of the brown dwarf cooling sequence. However, the observed color of the object is in disagreement with theoretical predictions for the chemical composition of \tuc{} \citep{roman_47Tuc}, suggesting that \textit{BD10} is likely a background star mistaken for a member of the cluster. The precise nature of \textit{BD10} can be determined by an additional visit of the field with JWST to verify cluster membership through proper motion.

Another NIRCam view of \tuc{} was later presented in \citet{marino_47tuc_BD} based on both new and archival \citep{JWST_proposal_2} observations. The inferred color-magnitude diagram (CMD) was in agreement with \citet{rolly_BD} down to the apparent end of the main sequence; however, unlike the CMD in \citet{rolly_BD}, the observed sequence in \citet{marino_47tuc_BD} appeared to extend to fainter magnitudes and redder colors just below an under-density of objects that \citet{marino_47tuc_BD} interpreted as a possible stellar/substellar gap. This candidate brown dwarf cooling sequence is also in better agreement with theoretical predictions, but clearly requires a more extensive grid of models with variable physical parameters to resolve quantitative discrepancies. As with \textit{BD10}, a second epoch of JWST photometry would provide a means to verify the nature of objects along the proposed substellar sequence and differentiate them from contaminants such as background galaxies. Detailed counts of \textit{bona fide} brown dwarfs verified in this manner are essential to use the substellar sequence in the determination of ages and mass functions across the hydrogen-burning limit.

This work is the second entry in a series of papers analyzing new NIRCam images of \ngc{}. The first paper in the series \citep{paperI} focuses on the white dwarf members of the cluster. This paper explores the brown dwarf cooling sequence and is organized as follows. Section~\ref{sec:observations} describes our observations of \ngc{} with JWST NIRCam, data reduction and cluster membership determination using archival HST images. Section~\ref{sec:isochrones} introduces a new grid of stellar models that we calculated for this analysis. Section~\ref{sec:BDs} presents the discovery of brown dwarfs in \ngc{} and derives their parameters. Section~\ref{sec:MF} details our determination of the age of the cluster from its brown dwarf cooling sequence. The mass function of the cluster near the hydrogen-burning limit is computed in the same section. Finally, our findings are summarized and conclusions are drawn in Section~\ref{sec:conclusion}.





\section{JWST observations} \label{sec:observations}
This study is based on the images of \ngc{} obtained with NIRCam on JWST\footnote{The data used in this paper is available in the Mikulski Archive for Space Telescopes (MAST), \dataset[doi: 10.17909/gxr7-dt60]{https://doi.org/10.17909/gxr7-dt60}} under program GO-1979 (PI: Bedin). The observed field covers a $\sim 2'\times8'$ patch of the sky, situated $\sim6'$ ($\sim 2$ half-light radii, \citealt{GC_distances}) from the center of the cluster. A complete description of the observed field, instrumental setup and data reduction can be found in the first paper of this series \citep{paperI}, and in \citet{JWST_phot_I,JWST_phot_II,JWST_phot_III,rolly_BD}. Here, we restrict ourselves to a brief summary of key details.

The observations were carried out simultaneously in the \texttt{F150W2} and \texttt{F322W2} bands of the \textit{Short Wavelength} and \textit{Long Wavelength} channels of NIRCam, respectively. These ultra-wide filters are particularly suitable for cooling brown dwarfs in globular clusters ($\teff\lesssim \qty{1500}{\K}$) by virtue of being centered near the two prominent peaks in the spectral energy distribution at $\sim\qty{1.5}{\micro\meter}$ and $\sim\qty{3.5}{\micro\meter}$, separated by the band of collision-induced $\ce{H_2}$ absorption at $\sim 2-\qty{3}{\micro\meter}$ \citep{CIA_0,CIA_2}. A total exposure time of $\sim\qty{2}{hr}$ was obtained across all pointings of the chosen dither pattern, allowing for usable photometry down to $>\qty{26}{mag}$\footnote{All magnitudes quoted in this paper are in the \texttt{VEGAMAG} system} in both filters, as required to capture the onset of the brown dwarf cooling sequence according to our earlier theoretical predictions \citep{roman_47Tuc}.

\begin{figure}[ht!]
    \centering
    \includegraphics[width=1\columnwidth]{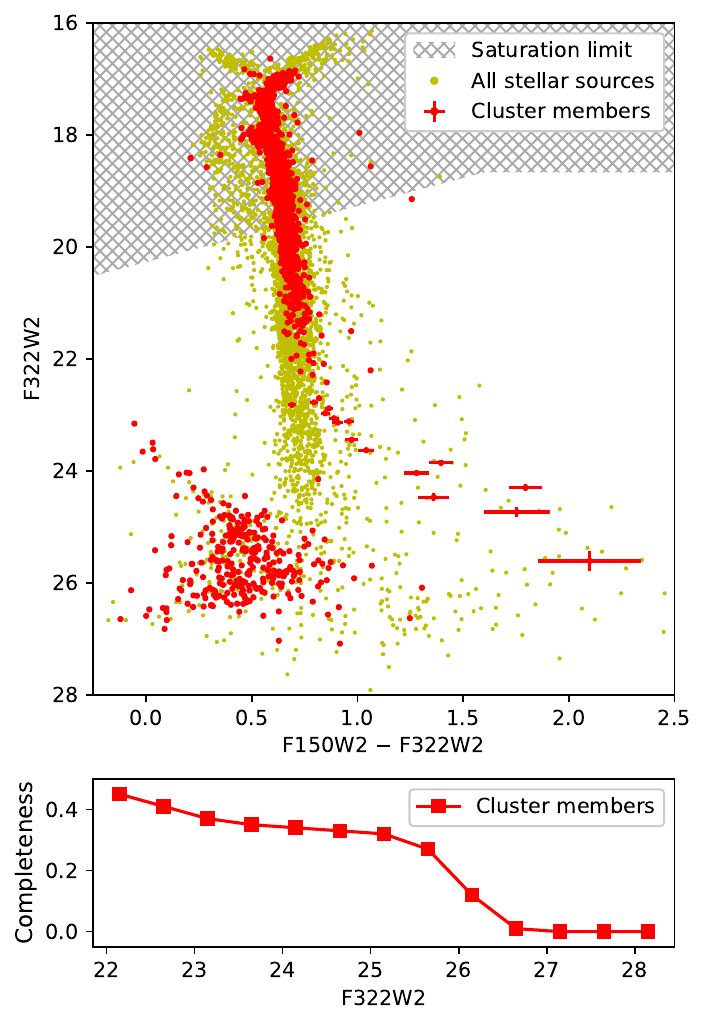}
    \caption{\textit{Top:} CMDs extracted from the JWST NIRCam observations in this study. Proper motion-selected members of \ngc{} are shown with red markers, while the rest of the stellar objects in the field are shown with yellow markers, including stars in the Galactic bulge and other non-members. The uncertainties in photometry derived from AS tests are shown for cluster members along the stellar/subtellar sequence. \textit{Bottom:} estimated completeness curve for the proper motion-cleaned CMD of \ngc{}, taken along the stellar/substellar sequence of the cluster.}
    \label{fig:CMD}
\end{figure}

The photometry of observed sources was extracted from the new JWST images using the multi-pass approach described in \citet{KS2_2}. The preliminary selection of stellar sources was carried out using the ``\textit{stellarity index}'' (\texttt{RADXS}, \citealt{RADXS}). The CMD of the identified stellar objects in the observed field is provided in Figure~\ref{fig:CMD} (yellow markers).

\ngc{} is situated at a low Galactic latitude ($b\approx-12^\circ$) and, hence, any field of observation in the vicinity of the cluster inevitably contains a large number of stars from the Galactic bulge. In Figure~\ref{fig:CMD}, the Galactic bulge forms a distinct sequence of background stars with a wide scatter in color and extending to fainter magnitudes than the end of the main sequence of \ngc{}. Additionally, some of the non-stellar sources may have passed the \texttt{RADXS} selection and contributed to further contamination. For the purposes of this study, we require a means to unambiguously identify the genuine members of \ngc{}, and remove the Galactic bulge, other non-members in the field and residual non-stellar sources.

We carried out our membership selection using proper motion measurements from multiple observation epochs. Since no other JWST observations of the target field are presently available, the first epoch was recovered from archival HST images. The field of our JWST observations was specifically chosen to overlap with the HST Advanced Camera for Surveys (ACS) visit under program GO-10424 (PI: Richer). \ngc{} was observed with ACS in the optical bands \texttt{F606W} and \texttt{F814W} over the course of 124 orbits, producing the deepest available optical photometry of any globular cluster. The temporal baseline between our JWST observations, and the archival HST data is approximately $18$ years (years 2023 vs 2005).

The internal velocity dispersion of \ngc{} (\qty{4.5}{\km/\s}, \citealt{GC_distances}) corresponds to the proper motion dispersion of $\sim\qty{0.4}{mas/yr}$ at the estimated distance of $\qty{2.458}{kpc}$ \citep{NGC6397_distance}. To accommodate cluster members with larger errors in the measured proper motions, we defined the \ngc{} membership criterion as $\leq\qty{2.5}{mas/yr}$ 
from the cluster average. The CMD of the objects that satisfy this criterion is shown in the upper panel of Figure~\ref{fig:CMD} with red markers. The proper motion-cleaned CMD clearly displays the end of the main sequence of \ngc{} around \texttt{F322W2} magnitude $\sim23$, followed by the rapid cooling (and reddening) of the members as their masses approach the hydrogen-burning limit. The white dwarf cooling sequence with \texttt{F322W2} magnitudes fainter than $\sim24$ and \texttt{F150W2}$-$\texttt{F322W2} colors bluer than $\sim1$ is also clearly visible, and is analyzed in detail in the first paper of this series.

To estimate the age of \ngc{} from the brown dwarf cooling sequence, the observed member counts as a function of magnitude must be corrected for objects missed by our final selection due to unreliable photometry, lack of identifiable counterparts in the HST images, and imperfect quality cuts. The completeness of our final CMD as a function of magnitude was estimated by injecting artificial stars (AS) \review{with zero proper motion} along the fiducial line of the stellar/substellar sequence, re-running the photometry and membership identification procedure, and determining the fraction of recovered AS, where an AS is considered recovered if its measured position and magnitude fall within $0.5$ pixels and $\qty{1}{mag}$ of the true values, respectively. The completeness curve as a function of the \texttt{F322W2} magnitude is shown in the lower panel of Figure~\ref{fig:CMD}.

\review{The AS tests were also used to determine the expected errors in our photometric and astrometric measurements along the stellar/substellar sequence as a function of magnitude. The proper motions of AS were best fit by the Rice distribution with the standard deviation of $\sim\qty{0.6}{mas/yr}$ down to the completeness limit at \texttt{F322W2} $\sim26.5$. We therefore expect reliable proper motion measurements for all stars that have passed our quality cuts and have detectable counterparts in archival HST images.}

\review{Using the best-fit theoretical isochrone (see Section~\ref{sec:isochrones}) and AS tests, we calculated the limiting magnitudes, down to which reliable JWST \texttt{F150W2}$-$\texttt{F322W2} and HST \texttt{F606W}$-$\texttt{F814W} colors can be obtained. Accurate photometric colors are necessary to differentiate brown dwarfs from white dwarfs and determine their stellar parameters. The JWST color errors derived from AS do not exceed $\qty{0.3}{mag}$ down to \texttt{F322W2}$\sim 25.5$. We therefore expect reliable JWST colors for all cluster members along the stellar/substellar sequence with the possible exception of the reddest member in the upper panel of Figure~\ref{fig:CMD}. On the other hand, the errors in HST colors are expected to exceed $\qty{0.3}{mag}$ at the equivalent \texttt{F322W2} JWST magnitude of $\sim 22$ (\texttt{F814W}$\sim25$, \texttt{F606W}$\sim29$). The limit is primarily driven by the low temperatures of stars near the hydrogen-burning limit that make them undetectable in the \texttt{F606W} band of ACS. Since astrometric measurements have a lower signal-to-noise threshold than photometry, and since they only require a single photometric band at each epoch, the archival HST data is suitable for accurate membership determination. However, our AS tests clearly indicate that JWST photometry is absolutely necessary to identify brown dwarf candidates and measure their stellar parameters.}

\section{Theoretical isochrones} \label{sec:isochrones}
\subsection{SANDee models}

To characterize the properties of brown dwarfs in \ngc{}, as well as other globular clusters in future photometric studies, we calculated a new grid of model isochrones and evolutionary tracks across the hydrogen-burning limit for a wide variety of chemical abundances, including metal-poor $\alpha$-enhanced compositions observed in globular clusters. The impact of atmospheres on stellar evolution (\textit{atmosphere-interior coupling}, \citealt{BC_origin}) is particularly prominent at low effective temperatures, and may offset theoretical CMDs by multiple magnitudes \citep{sky_47_tuc}. To account for atmosphere-interior coupling in the new isochrones, we used the temperature-pressure profiles from the \texttt{SAND} (\texttt{S}pectral \texttt{AN}alog of \texttt{D}warfs) grid of low-temperature model atmospheres \citep{efrain} as the outer boundary conditions in the equations of stellar structure. We refer to the new grid of model isochrones as \texttt{SANDee} (\texttt{e}volutionary \texttt{e}xtension to \texttt{SAND}). \texttt{SAND} atmospheres (and, by extension, the new \texttt{SANDee} isochrones) include $24$ different chemistries parameterized by metallicity ($\ce{[Fe/H]}$) and $\alpha$-enhancement ($\ce{[\alpha/Fe]}$). The specific values provided by \texttt{SAND} were chosen to uniformly cover the $\ce{[Fe/H]}-\ce{[\alpha/Fe]}$ distribution of the Milky Way inferred from the APOGEE survey \citep{APOGEE,APOGEE_DR14,APOGEE_chem_distribution}. The available metallicities in the \texttt{SANDee} grid range from $\ce{[Fe/H]}=-2.4$ to $+0.3$.

\texttt{SANDee} isochrones are based on new evolutionary models computed with the \texttt{MESA} (\texttt{M}odules for \texttt{E}xperiments in \texttt{S}tellar \texttt{A}strophysics) code, version \texttt{23.05.1} \citep{MESA,MESA_2,MESA_3,MESA_4,MESA_5}. Minor changes to the \texttt{MESA} codebase were introduced to ensure compatibility with \texttt{SAND} atmospheres, following our previous work \citep{roman_47Tuc}. All models and the patch file with our customizations are made publicly available in our online model repository\footnote[1]{\href{https://romanger.com/models.html}{https://romanger.com/models.html}} and on Zenodo\footnote{\dataset[https://doi.org/10.5281/zenodo.10989779]{https://doi.org/10.5281/zenodo.10989779}}.

For each chemistry, the initial stellar masses of the evolutionary models were sampled adaptively to ensure that the difference between the luminosities and effective temperatures of adjacent models did not exceed $\qty{0.12}{dex}$ and $\qty{120}{\K}$ respectively, at all ages between $\qty{1.0}{Gyr}$ and $\qty{13.5}{Gyr}$ in steps of $\qty{0.5}{Gyr}$. The evolution was run from the pre-main sequence phase until the age of $\qty{13.5}{Gyr}$, or until the model evolved beyond the temperature range of \texttt{SAND} atmospheres ($\lesssim\qty{1000}{\K}-\qty{4000}{\K}$ with the specific lower bound dependent on chemistry). The lowest mass for each isochrone was set to $\qty{0.06}{\Msun}$ -- a value well below the hydrogen-burning limit for all considered chemistries -- while the highest mass was set to match the point at which the model begins to exceed $\qty{4000}{\K}$ in effective temperature before the age of $\qty{1.0}{Gyr}$. This upper limit varies between $\sim\qty{0.15}{\Msun}$ at the lowest considered metallicities and $\sim\qty{0.6}{\Msun}$ at the highest.

Unlike our previous work \citep{roman_47Tuc} where multiple atmosphere-interior coupling schemes were blended together, we calculated all \texttt{SANDee} models with atmosphere boundary conditions at the Rosseland mean optical depth of $100$ (see discussion in \citealt{tau_100_vs_photosphere}). The majority of models converged within the ``\textit{gold tolerances}'' \citep{MESA_5} at every time step. For a small number of models, the gold tolerances had to be relaxed to complete the computation. A number of \texttt{SAND} model atmospheres were excluded from the atmosphere-interior coupling calculation to avoid poor convergence due to discontinuities in the outer boundary conditions. A readme file is provided alongside the models that details all accommodations for poor convergence made in \texttt{SANDee}.

\begin{figure}[ht!]
    \centering
    \includegraphics[width=1\columnwidth]{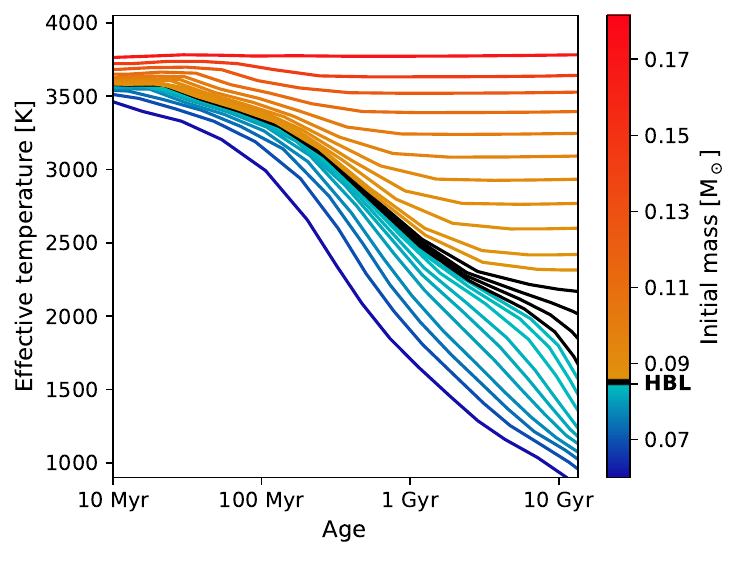}
    \caption{Cooling curves for stellar and substellar objects with initial masses between $\qty{0.06}{\Msun}$ and $\sim\qty{0.18}{\Msun}$, evaluated from the \texttt{SANDee} model isochrones for $\ce{[Fe/H]}=-1.75$ and $\ce{[\alpha/Fe]}=0.3$. The hydrogen-burning limit (HBL) divides the objects into main sequence stars that attain a steady state, and brown dwarfs that continue cooling long-term.}
    \label{fig:cooling_curves}
\end{figure}

To demonstrate the application of \texttt{SANDee} models, a set of theoretical cooling curves for one of the considered chemistries ($\ce{[Fe/H]}=-1.75$, $\ce{[\alpha/Fe]}=0.3$) is plotted in Figure~\ref{fig:cooling_curves}. We make the \texttt{SANDee} grid publicly available alongside a Python script that allows the user to compute model CMDs for any combination of photometric bands and reddening parameters.

\subsection{Hydrogen-burning limit}

Since the boundary between main sequence stars and brown dwarfs is central to this study, we examined the new \texttt{SANDee} models to develop a method for the unambiguous identification of the hydrogen-burning limit in observed CMDs. Choosing a singular stellar mass as the hydrogen-burning limit is not straightforward, since $L_\mathrm{nuc}/L$ (where $L$ is the total luminosity output of the star, and $L_\mathrm{nuc}$ is the energy production rate by nuclear fusion) is a continuously varying quantity that evolves with stellar age.

\begin{figure}[ht!]
    \centering
    \includegraphics[width=1\columnwidth]{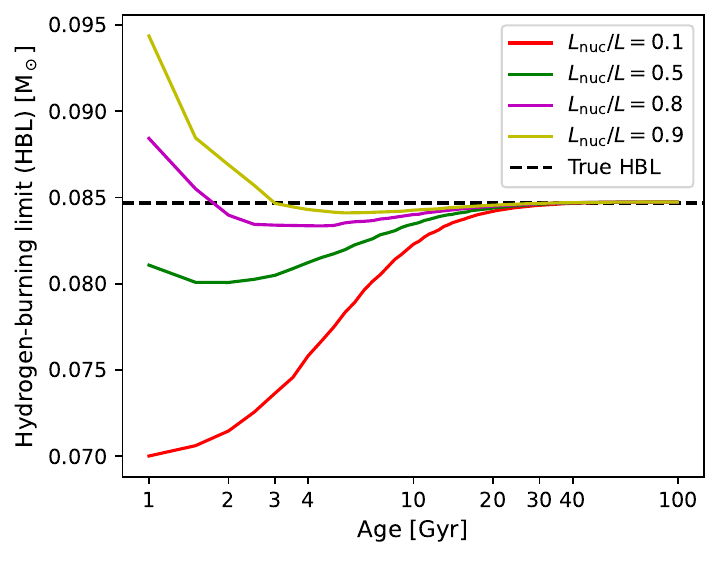}
    \caption{Dependence of the hydrogen-burning limit on the value of $L_\mathrm{nuc}/L$ adopted as the boundary between brown dwarfs and main sequence stars, and the stellar age at which $L_\mathrm{nuc}/L$ is considered. The data in the figure were derived by extending the evolutionary models of one of the \texttt{SANDee} isochrones ($\ce{[Fe/H]}=-1.75$, $\ce{[\alpha/Fe]}=0.3$) to \qty{100}{Gyr}. The true hydrogen-burning limit indicated with the dashed line was computed using the definition in text.}
    \label{fig:HBL}
\end{figure}

Figure~\ref{fig:HBL} illustrates how the definition of the hydrogen-burning limit varies with stellar age and the specific choice of $L_\mathrm{nuc}/L$ as the boundary between main sequence stars and brown dwarfs. The figure demonstrates that even at the age of the Universe, the value of the hydrogen-burning limit is significantly affected by the adopted $L_\mathrm{nuc}/L$ threshold. Therefore, for all populations of stellar objects in the present-day Universe there exists a range of initial stellar masses on the order of $\Delta M\gtrsim\qty{0.003}{\Msun}$ (which may translate into multiple apparent magnitudes due to the rapid evolution of objects near the hydrogen-burning limit, see Section~\ref{sec:MF}), corresponding to yet ``undecided'' objects with $0<L_\mathrm{nuc}/L<1$. These objects may evolve into brown dwarfs or main sequence stars in the future, eventually attaining $L_\mathrm{nuc}/L\approx0$ or $L_\mathrm{nuc}/L\approx1$ respectively. However, at the age of $\qty{100}{Gyr}$ the definition of the hydrogen-burning limit is largely insensitive to $L_\mathrm{nuc}/L$, as nearly all ``undecided'' objects will have had enough time to settle as unambiguous brown dwarfs or stars.

\begin{figure}[ht!]
    \centering
    \includegraphics[width=1\columnwidth]{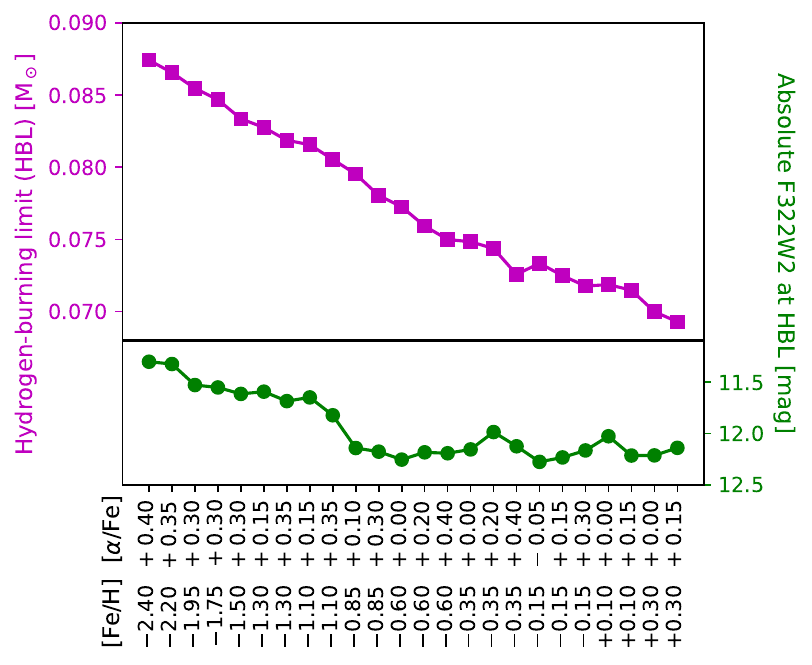}
    \caption{True hydrogen-burning limits for all $24$ chemistries available in \texttt{SANDee} (magenta, left vertical axis), and the corresponding absolute NIRCam \texttt{F322W2} magnitudes at the true hydrogen-burning limit (green, right vertical axis). The horizontal axis is categorical, ordered first by $\ce{[Fe/H]}$, then by $\ce{[\alpha/Fe]}$. Magnitudes are calculated at $\qty{13}{Gyr}$.}
    \label{fig:HBL_chem}
\end{figure}

For the purposes of this study, we propose a definition of the \textit{true hydrogen-burning limit}, obtained by determining the initial stellar mass of the object that attains $L_\mathrm{nuc}/L=0.5$ at the age of $\qty{100}{Gyr}$. To derive the true limit for every \texttt{SANDee} chemistry, we calculated a set of additional \texttt{MESA} models to $\qty{100}{Gyr}$, specifically searching for the initial stellar mass that ensures the above condition using Brent's method with the absolute tolerance of \qty{e-4}{\Msun}. The calculated limits are plotted in Figure~\ref{fig:HBL_chem} for all chemistries available in \texttt{SANDee} both as masses and absolute magnitudes. The figure demonstrates that both $\ce{[Fe/H]}$ and $\ce{[\alpha/Fe]}$ may offset the hydrogen-burning limit by $\sim0.5-\qty{1}{mag}$.

%
%

We close this section by evaluating the theoretical hydrogen-burning limits derived from \texttt{SANDee} against the only reported observation of the candidate stellar/substellar gap in a globular cluster made by \citet{marino_47tuc_BD}. The local minimum in the luminosity function of \tuc{} derived by the authors falls near the \texttt{F322W2} apparent magnitude $25.3$. The closest \texttt{SANDee} chemistries to the spectroscopic metallicity of \tuc{} ($\ce{[Fe/H]}\approx-0.75$, \citealt{nominal_C14,nominal_M16,nominal_T14,roman_47Tuc}) are the two $\ce{[Fe/H]}=-0.85$ models (see Figure~\ref{fig:HBL_chem}) that both predict the hydrogen-burning limit magnitude within $\qty{0.1}{mag}$ of the observation, using the distance modulus of $13.2$ \citep{dm}. On the other hand, our earlier prediction of the stellar/substellar gap in \tuc{} \citep{roman_47Tuc} appears to place it $\sim0.5$ magnitudes fainter, despite accounting for the detailed distribution of chemical abundances in the cluster that is completely ignored in \texttt{SANDee}.

The reason for the apparent discrepancy between these predictions is likely two-fold. First, the local minimum in the luminosity function may not exactly correspond to the hydrogen-burning limit due to the non-straightforward mass-luminosity relationship. Second, the atmospheric chemistry of low-temperature members may vary with stellar mass, resulting in a measurable difference between the chemical abundances inferred from higher-temperature stars, and the best-fit chemical abundances at the hydrogen-burning limit. The offset between the stellar/substellar gap and the hydrogen-burning limit is calculated in Section~\ref{sec:MF}. The dependence of chemistry on stellar mass is examined next.

\section{Brown dwarfs in NGC~6397} \label{sec:BDs}
\subsection{CMD fitting}

To infer the properties of brown dwarfs in \ngc{}, we first identified the chemical composition from the \texttt{SANDee} grid that provides the best approximation to the observed CMD of the cluster. \texttt{SANDee} isochrones were converted to the color-magnitude space using the \texttt{SAND} model atmospheres at the corresponding chemistry, following \citet{roman_omega_cen} and \citet{mikaela}. To evaluate synthetic photometry, we adopted the optical reddening of $E(B-V)=\qty{0.22}{mag}$ from \citet{matteo_NGC6397}, the distance of $\qty{2.458}{kpc}$ from \citet{NGC6397_distance}, the extinction law from \citet{extinction} and the total-to-selective extinction ratio of $R_V=3.1$. For CMD fitting, we evaluated all isochrones at the age of $\qty{12.6}{Gyr}$ \citep{matteo_NGC6397}; however, the color-magnitude relation predicted by the isochrone is largely insensitive to age in the observed magnitude range, and therefore the specifics of the latter choice have little importance.

\begin{figure}[ht!]
    \centering
    \includegraphics[width=1\columnwidth]{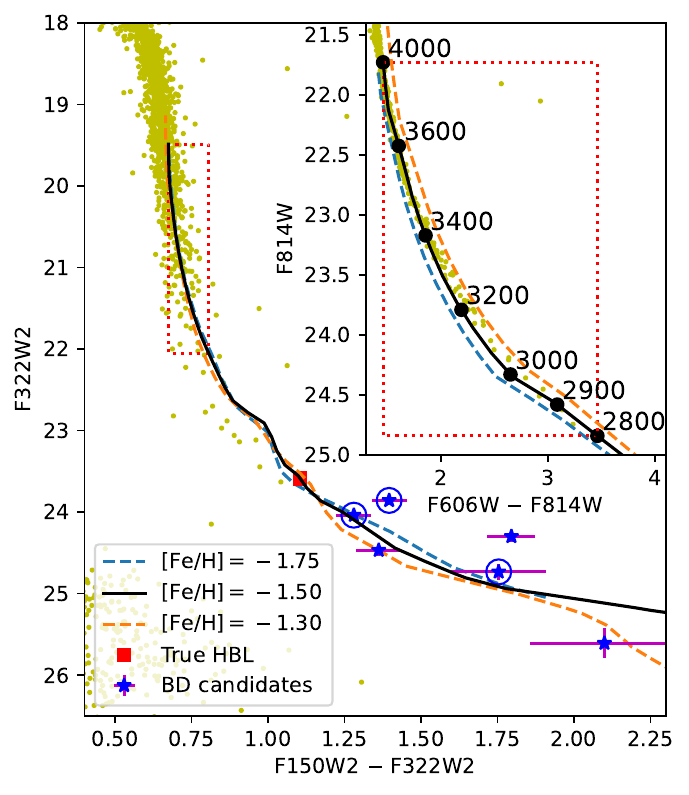}
    \caption{Proper motion-cleaned CMD of \ngc{} as observed with JWST NIRCam. This CMD corresponds to the red markers in Figure~\ref{fig:CMD}. Three best-fitting \texttt{SANDee} isochrones are overplotted, of which the $\ce{[Fe/H]}=-1.5$ isochrone (black) was used in our analysis. The true hydrogen-burning limit (HBL), as defined in Section~\ref{sec:isochrones}, is indicated for that isochrone with the red marker. The brown dwarf candidates in the cluster are highlighted with blue markers. The candidates confirmed by visual inspection are circled. The photometric uncertainties for the brown dwarf candidates were derived from AS tests as detailed in Section~\ref{sec:observations}. The inset shows a cutout of the HST ACS CMD with the same three isochrones overplotted, and selected effective temperatures on the $\ce{[Fe/H]}=-1.5$ isochrone indicated with black markers and labelled in Kelvin. The range of stellar parameters contained within the red dotted rectangle is approximately the same in both CMDs.}
    \label{fig:best_fit_iso}
\end{figure}

The three best-fitting \texttt{SANDee} isochrones are shown in Figure~\ref{fig:best_fit_iso} for both the new JWST NIRCam data described in Section~\ref{sec:observations} and a subset of archival HST ACS observations, for which usable photometry and model coverage were available. All three isochrones are $\alpha$-enhanced ($\ce{[\alpha/Fe]}=0.35$ for the $\ce{[Fe/H]=-1.3}$ isochrone, and $\ce{[\alpha/Fe]}=0.3$ for the other two). Of the three candidate isochrones, $\ce{[Fe/H]}=-1.5$ appears to provide the best fit, as determined primarily by the HST CMD. On the other hand, the JWST CMD is largely insensitive to metallicity within the available magnitude range above the hydrogen-burning limit, and the photometric scatter below the hydrogen-burning limit far exceeds the differences between the shown isochrones.

It is somewhat puzzling that despite the excellent agreement with both HST and JWST photometry, the best-fit isochrone suggests the cluster metallicity of $\ce{[Fe/H]}=-1.5$, which is significantly larger than the photometric (e.g. $\ce{[Fe/H]}=-1.88\pm0.04$ in \citealt{matteo_NGC6397}) and spectroscopic (e.g. $\ce{[Fe/H]}=-2.120\pm0.002$ in \citealt{NGC6397_MUSE}) estimates in the literature. It must be emphasized that our analysis was carried out on the lower main sequence near the hydrogen-burning limit, while all previous measurements were largely derived from upper main sequence and post-main sequence members. The HST CMD in Figure~\ref{fig:best_fit_iso}, in fact, indicates that the best-fit metallicity may be decreasing towards higher stellar masses across the magnitude range shown. While all members of the cluster are expected to share approximately the same true metallicity, the effective metallicity estimated from CMD fitting may vary at lower masses due to imperfectly modelled low-temperature effects, such as condensation of gaseous species, gravitational settling, non-equilibrium chemistry, and variations in molecular chemistry and opacity due to the abundances of individual elements. 

The gradual drift of best-fit chemistry as a function of metallicity is the probable cause of the discrepancy between the predicted luminosity of the stellar/substellar gap in \tuc{} from \citet{roman_47Tuc}, and its much brighter alternative presented in this work. In particular, the predictions derived in \citet{roman_47Tuc} were based on the assumption that the majority of cluster members are oxygen-enhanced. On the other hand, the apparent agreement between \citet{marino_47tuc_BD} and the $\ce{[Fe/H]}=-0.85$ \texttt{SANDee} model with solar oxygen suggests that the effective value of $\ce{[O/Fe]}$ near the hydrogen-burning limit may be significantly lower than observed in higher-$\teff{}$ stars, e.g. due to more efficient condensation of oxygen-bearing species onto dust grains than predicted by model atmospheres. This effect may also be related to the known discrepancies in atmospheric chemical abundances in star / brown dwarf binary systems in the field \citep{BD_inconsistent_chemistry}. Globular clusters provide a unique opportunity to study the gradual emergence of this effect with decreasing stellar mass.

Since the $\ce{[Fe/H]}=-1.5$ \texttt{SANDee} isochrone provides the best fit to the observations, we used it in the subsequent analysis.

\subsection{Brown dwarf candidates}

We consider observed sources in our JWST NIRCam images to be \textit{bona fide} brown dwarf members of \ngc{} if and only if they satisfy all of the following criteria:

\begin{enumerate}
    \item The source must pass the ``stellarity index'' (\texttt{RADXS}) test \citep{RADXS}.
    \item The source must have an identifiable counterpart in the archival HST images, and the measured proper motion must be consistent with the average velocity and velocity dispersion of \ngc{}, as detailed in Section~\ref{sec:observations}.
    \item The presence of the source in both HST and JWST images at positions consistent with its inferred proper motion must be visually confirmed by a human.
    \item The source must have \texttt{F322W2} and \texttt{F150W2} magnitudes consistent with the substellar color-magnitude relation predicted by the best-fit \texttt{SANDee} isochrone.
\end{enumerate}

We identified preliminary brown dwarf candidates in the proper motion-cleaned JWST CMD of \ngc{} (Figure~\ref{fig:best_fit_iso}) by selecting the members that have redder colors than the color at the true hydrogen-burning limit predicted by the isochrone (Figure~\ref{fig:HBL_chem}), and magnitudes within $\qty{1}{mag}$ of the predicted color-magnitude trend. The six suitable brown dwarf candidates are highlighted in Figure~\ref{fig:best_fit_iso} with blue markers. We note that while all six candidates have HST cross-matches with proper motions suggestive of cluster membership, they do not have reliable HST photometry due to their exceedingly faint optical magnitudes, \review{especially in the \texttt{F606W} band}.

\begin{figure*}
    \centering
    \includegraphics[width=1.2\columnwidth]{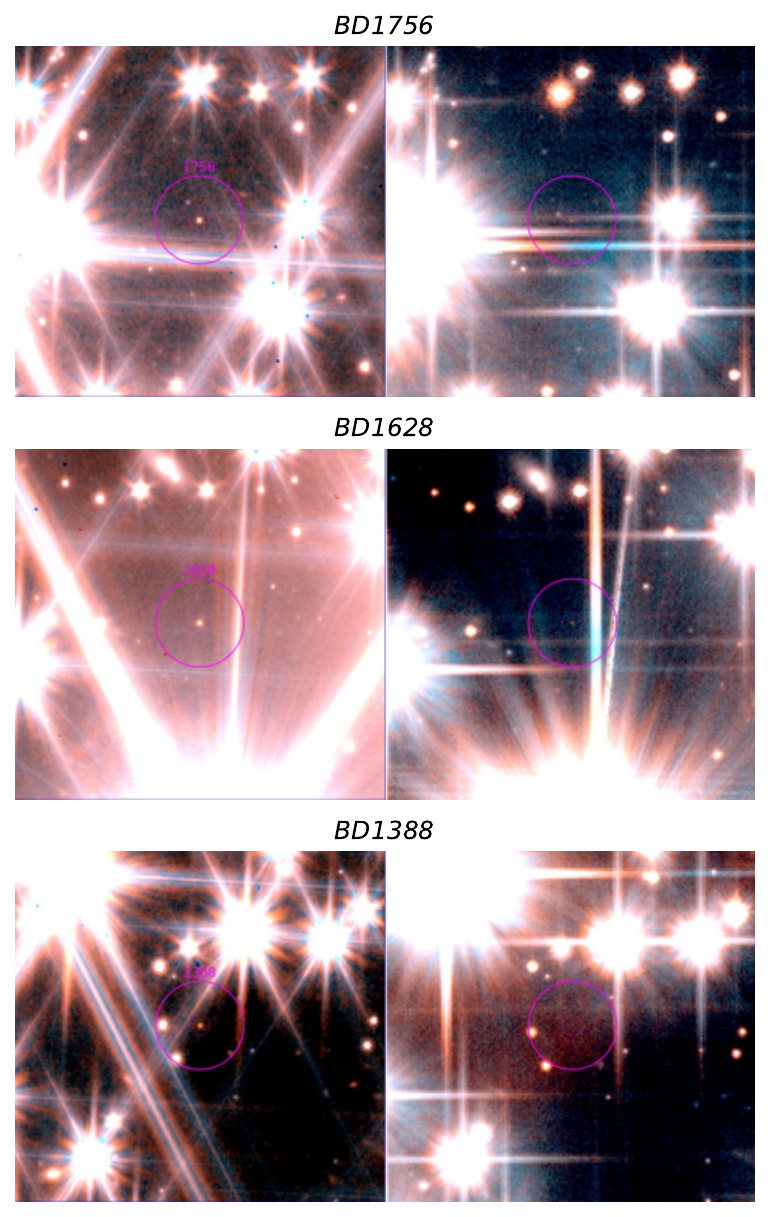}
    \caption{Cutouts of JWST NIRCam (\textit{left}) and HST ACS (\textit{right}) images of \ngc{}, centered on the three \textit{bona fide} brown dwarf members of the globular cluster. All images are RGB with the red channel representing \texttt{F322W2}/\texttt{F814W}, blue channel representing \texttt{F150W2}/\texttt{F606W}, and green channel populated with a linear combination of the other two channels for clarity. The overplotted annuli in the JWST images are placed over the measured coordinates of the objects. Their counterparts in the HST images are placed over the coordinates, converted to the HST epoch using the measured proper motions.}
    \label{fig:finders}
\end{figure*}

We inspected all six candidates visually in both JWST and HST images. Of the six candidates, only three (hereby designated as \textit{BD1756}, \textit{BD1628} and \textit{BD1388}) appeared as unambiguous sources in both JWST and HST images, at the coordinates consistent with the measured proper motion. We therefore only treat these three objects as \textit{bona fide} brown dwarf members of \ngc{}. The relevant cutouts of the images with marked coordinates of the brown dwarf members are provided in Figure~\ref{fig:finders}. The confirmed candidates are circled in Figure~\ref{fig:best_fit_iso}.

\subsection{Stellar parameters}

In this subsection, we use the terms \textit{color} and \textit{magnitude} to specifically refer to JWST NIRCam \texttt{F150W2}$-$\texttt{F322W2} colors and \texttt{F322W2} magnitudes, as in Figure~\ref{fig:best_fit_iso}. The stellar parameters of the confirmed brown dwarf members were derived by comparing their measured colors and magnitudes with the predictions of the model isochrone. The small sample size of brown dwarf members, and the lack of reliable estimates of the systematic errors in \texttt{SANDee} models make the error analysis for the derived parameters particularly challenging. However, preliminary results may be obtained using reasonable simplifying assumptions described below.

The average error in the observed brown dwarf magnitudes was estimated as the population standard deviation of the scatter around the best-fit isochrone:

\begin{equation}
    \sigma_\mathrm{m}=\sqrt{\frac{n}{n-1}\left(\langle [m-m_\mathrm{iso}(c)]^2 \rangle - \langle m-m_\mathrm{iso}(c) \rangle^2\right)}
    \label{eq:mag_error}
\end{equation}

Here, $n$ is the number of \textit{bona fide} brown dwarf members ($3$), $\sigma_m$ is the average error in the measured brown dwarf magnitudes, $m$ is the set of measured brown dwarf magnitudes, $c$ is the set of their measured colors, and $m_\mathrm{iso}$ is the color-magnitude relationship predicted by the best-fit isochrone. The scatter was calculated as $\sigma_m\approx \qty{0.25}{mag}$. We note that Equation~\ref{eq:mag_error} overestimates the true value of $\sigma_m$, since the scatter of measured magnitudes around the isochrone is included in the equation twice: directly as the scatter in $m$, and indirectly as the scatter in $c$. Since the fractional contributions of these effects are not known, we chose to leave this conservative estimate of $\sigma_m$ uncorrected; however, the real value of $\sigma_m$ may be up to a factor of $\sqrt{2}$ lower. The value of $\sigma_m$ was used as an estimate of the magnitude error for all brown dwarf members.

The masses of brown dwarfs, $M$, were estimated by interpolating the best-fit isochrone at their measured colors, $c$, since $c$ is far more sensitive to $M$ in the brown dwarf regime than $m$, as seen in Figure~\ref{fig:best_fit_iso}. The errors in the inferred masses were estimated using the mass-magnitude relationship provided by the isochrone, and the value of $\sigma_m$. Our approach to estimate masses from the observed colors, and mass errors from the observed magnitudes implies the assumption that the errors in colors and magnitudes are related by the derivative of the theoretical color-magnitude relationship ($d(m_\mathrm{iso})/d(c_\mathrm{iso})$). This assumption is expected to hold if the photometric errors are largely systematic (e.g. due to unaccounted abundances of individual elements), which should be the case given that the magnitude limit of JWST photometry falls far below the observed brown dwarf sequence.

Finally, the rest of the stellar parameters of interest (effective temperature, $\teff$; luminosity, $L$; surface gravity, $\logg$; stellar radius, $R$) were estimated from the calculated masses and mass errors, assuming that all errors are Gaussian and the parameter relationships derived from the best-fit isochrone are approximately linear within the magnitude of said errors. More accurate errors (including asymmetric error bars) can be estimated by sampling the parameter distribution with Monte Carlo methods; however, given the preliminary nature of our approach to estimating the errors in colors and magnitudes, we deemed the added benefit of a more involved error analysis insignificant.

The estimated stellar parameters of the three \textit{bona fide} brown dwarfs and their errors are listed in Table~\ref{tab:bd_params}, alongside their apparent NIRCam magnitudes and measured celestial coordinates.

\startlongtable
\begin{deluxetable*}{l|ccccccc}
\tablecaption{Stellar parameters of the identified brown dwarf members in \ngc{} \label{tab:bd_params}}
\tablewidth{\textwidth}
\tablehead{
\colhead{Identifier} & Coordinates & \texttt{F150W2} / \texttt{F322W2} & $M/M_\mathrm{jup}$ & $\teff$ $[\mathrm{K}]$ &  $\log_{10}(L/L_\odot)$ & $\logg$  & $R/R_\mathrm{jup}$
}
\startdata
\textit{BD1756} & $17$:$41$:$11.6$ $-53$:$42$:$58.0$ & $25.32$ / $24.04$ & $86.7\pm 0.2$ & $1756\pm 111$ & $-4.27\pm 0.13$ & $5.56\pm 0.02$ & $0.771\pm 0.016$ \\
\textit{BD1628} & $17$:$41$:$08.8$ $-53$:$45$:$56.0$ & $25.25$ / $23.86$ & $86.5\pm 0.3$ & $1629\pm 102$ & $-4.42\pm 0.12$ & $5.58\pm 0.01$ & $0.753\pm 0.014$ \\
\textit{BD1388} & $17$:$40$:$58.1$ $-53$:$43$:$13.2$ & $26.49$ / $24.73$ & $85.6\pm 0.7$ & $1388\pm 124$ & $-4.74\pm 0.17$ & $5.61\pm 0.02$ & $0.721\pm 0.017$ \\
\enddata

\tablecomments{
The coordinates are provided in the Gaia DR3 \citep{Gaia_DR3} reference frame at the epoch of NIRCam observations (J2023.198063).}

\end{deluxetable*}

\section{Age and mass function of NGC~6397} \label{sec:MF}
In this section, we estimate the age of \ngc{} from its brown dwarf cooling sequence, as well as the mass function of the cluster across the hydrogen-burning limit. The CMD of main sequence stars in globular clusters is largely insensitive to the cluster age due to the slowly evolving steady state, attained by low-mass hydrogen-burning members. However, stellar members just above the hydrogen-burning limit (i.e. the ``undecided'' objects discussed in Section~\ref{sec:isochrones}) may evolve faster as they have not yet reached the energy equilibrium of the main sequence. Brown dwarf members below the hydrogen-burning limit are expected to evolve rapidly due to continuous cooling, as seen in Figure~\ref{fig:cooling_curves}. We therefore expect the luminosity function of \ngc{} near the hydrogen-burning limit and below it to serve as an age diagnostic. Even a small number of confirmed brown dwarf members may be sufficient to establish an upper limit on the age of the cluster, as the mere fact that brown dwarf members can be seen implies that \ngc{} is not sufficiently old to have allowed them to cool beyond the limit of the observed CMD. A more detailed discussion of brown dwarfs' potential as globular cluster ``clocks'' is available in our previous theoretical investigation \citep{roman_47Tuc}. The unprecedented photometric sensitivity of JWST for the first time allows this potential to be realized in practice.

We computed the observed luminosity function of \ngc{} by counting the number of observed stellar members and brown dwarfs in $10$ uniform \texttt{F322W2} magnitude bins between $m=19.7$ and $25$. 
The bright limit of the magnitude range is determined by both the upper mass limit of the \texttt{SANDee} grid and the saturation limit of our observations, while the faint limit is set by the magnitude of the lowest-mass brown dwarf observed in the cluster (\textit{BD1388}). Our star counts include all proper-motion confirmed members of \ngc{} above the true hydrogen-burning limit, and only the three \textit{bona fide} brown dwarf members below the hydrogen-burning limit.

\begin{figure}[ht!]
    \centering
    \includegraphics[width=1\columnwidth]{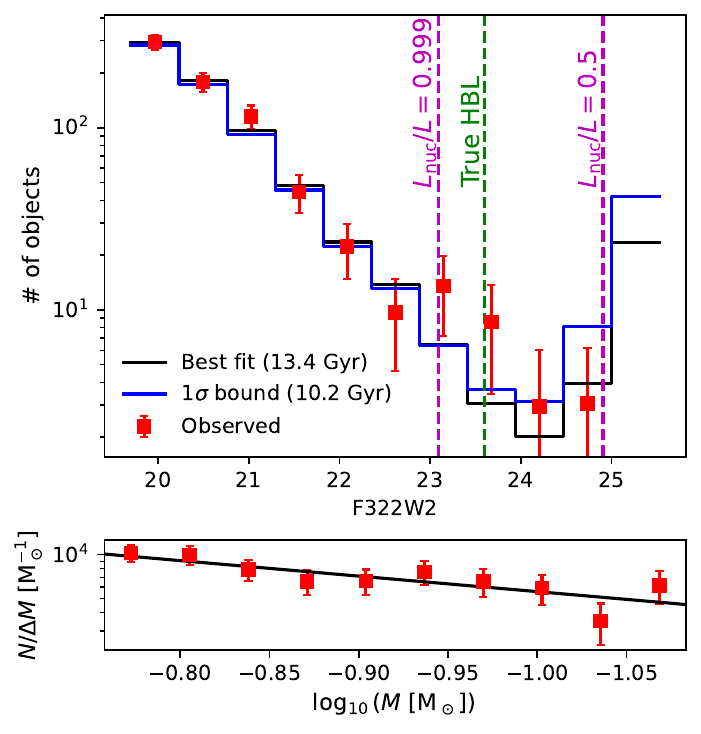}
    \caption{\textit{Top:} observed luminosity function of \ngc{}, overplotted on two theoretical models, corresponding to the best-fit values of $\alpha$ and age, as well as the best-fit value of $\alpha$ and the lower $1$-sigma bound on age. The true hydrogen-burning limit, as defined in Section~\ref{sec:isochrones}, as well as two alternative choices for $L_\mathrm{nuc}/L$ at the best-fit age are shown for reference. \textit{Bottom:} inferred mass function of \ngc{}, overplotted on the best-fit unbroken power law model, using the model isochrone at the best-fit age. Here $N$ is the object count in the bin, and $\Delta M$ is the width of the bin. In both panels, the displayed object counts have been corrected for the photometric completeness.}
    \label{fig:LF}
\end{figure}

The counts in each bin were corrected by the estimated photometric completeness from the lower panel of Figure~\ref{fig:CMD}. The errors in the counts were calculated by adding in quadrature the Poisson contribution due to the chosen field of observation, and the binomial contribution due to incomplete selection:

\begin{equation}
    \sigma_N=\sqrt{\left(\sqrt{N P (1-P)}\right)^2 + \left(\frac{\sqrt{N P}}{P}\right)^2}
    \label{eq:LF_errors}
\end{equation}

\noindent where $\sigma_N$ is the count error in a given bin, $N$ is the completeness-corrected object count in the bin, and $P$ is the estimated completeness fraction in the bin. The completeness-corrected observed luminosity function is plotted in the upper panel of Figure~\ref{fig:LF} with red markers.

The theoretical luminosity function can be estimated from the mass-luminosity relationship provided by the best-fit isochrone, and the mass function of the cluster. The luminosity function models used in our previous studies of the globular clusters \tuc{} \citep{roman_47Tuc} and \omegacen{} \citep{roman_omega_cen} determined that the mass functions of globular clusters can be accurately represented down to the hydrogen-burning limit as broken power laws. However, due to the narrower range of stellar masses considered in this study and the smaller sample size, we chose to adopt the unbroken power law instead:

\begin{equation}
    \xi(M)\propto M^{-\alpha}
    \label{eq:powerlaw}
\end{equation}

\noindent where $\xi(M)$ is the mass function, and $\alpha$ is the (negative) power law index. The theoretical luminosity function ($\phi$) can then be estimated as follows:

\begin{equation}
    \phi(m)=-\xi\left(M_\mathrm{iso}(m)\right) \frac{dM_\mathrm{iso}(m)}{dm_\mathrm{iso}}
    \label{eq:MF_to_LF}
\end{equation}

\noindent where $M_\mathrm{iso}(m)$ is the mass-luminosity relation provided by the best-fit isochrone, and $dM_\mathrm{iso}/dm_\mathrm{iso}$ is its derivative, evaluated numerically using finite differences.

In our analysis, we treated both the age of the best-fit isochrone and the power law index ($\alpha$) as free parameters, and determined their values by fitting the theoretical luminosity function integrated in the appropriate bins onto the observed luminosity function using the weighted linear least squares regression. The best-fit values and their errors were calculated as follows:

%
\begin{equation}
    \alpha=-1.06\pm0.22\ \ \ \ \mathrm{age}=13.4\pm3.3\ \qty{}{Gyr}
    \label{eq:results}
\end{equation}

The best-fit theoretical luminosity function is shown in the upper panel of Figure~\ref{fig:LF} in black. An alternative model corresponding to the $1$-sigma lower limit on the age of \ngc{} is shown in blue.

Since we found $\alpha<0$, the observed luminosity function is consistent with a top-heavy mass function. The ``universal initial mass function'' \citep{Kroupa}, on the other hand, suggests a bottom-heavy mass function in all regimes, but with a highly uncertain power index near the hydrogen-burning limit of $\alpha=0.3\pm0.7$. \citet{NGC6397_MF_1} estimated a top-heavy value of $\alpha=-0.4\pm0.4$ in the low-mass regime of \ngc{}, while \citet{NGC6397_MF_2}\footnote{Note that \citet{NGC6397_MF_2} use the opposite sign convention for $\alpha$, compared to \citet{Kroupa} and this work.} calculated $\alpha=0.4\pm0.3$, in agreement with the universal initial mass function. Finally, the mass function of \ngc{} inferred in \citet[][lower panel of their Figure~4]{ngc_6397_king} appears bottom-heavy at high stellar masses ($\gtrsim\qty{0.25}{\Msun}$) with $\alpha\sim 1$, but flattens out at lower masses ($\alpha\sim 0$), and shows a steep top-heavy drop-off at the hydrogen-burning limit with $\alpha<<-1$, likely accentuated by small-number statistics and the uncertain mass-luminosity relationship adopted in the low-mass regime.

The large spread in inferred mass functions within the same cluster emphasizes that globular clusters are dynamically complex systems, and the distribution of stellar masses may vary greatly with the choice of field and considered mass range. The value of $\alpha$ calculated in this study is not directly comparable to any of the literature values, since our estimate is derived from objects of lower masses and over a narrower range ($\qty{0.08}{\Msun}<M<\qty{0.18}{\Msun}$). There is also little reason to expect consistency between the local present-day mass function of the cluster, and the global initial mass function, even if the latter can be well-approximated by some universal form. 


If our analysis is restricted to main sequence stars only (i.e., if the four rightmost bins in the observed luminosity function in Figure~\ref{fig:LF} are excluded), $\alpha=-0.5\pm0.7$ is obtained, consistent with both flat and bottom-heavy mass functions within its error bars.

%
Our estimate of the age of \ngc{} from the brown dwarf cooling sequence is, in general, not sensitive to the adopted mass function. The best-fit value increases by less than $\qty{1}{Gyr}$ if the bottom-heavy mass function from \citet{Kroupa} with $\alpha=0.3$ is adopted instead of the best-fit value of $\alpha$. The effect is contained well within the quoted uncertainty in age. Likewise, fixing the cluster age to the main sequence turnoff value of $\qty{12.6}{Gyr}$ \citep{matteo_NGC6397} has an insignificant effect on the inferred mass function, offsetting the best-fit $\alpha$ by $\sim 0.01$.

The inferred mass function of \ngc{} and the best-fit power law model are plotted in the lower panel of Figure~\ref{fig:LF}. The mass function was calculated in $10$ log-uniform bins within the range of masses that corresponds to the range of magnitudes used in the derived luminosity function ($19.7<m<25$). The counts in each bin are completeness-corrected, and the errors were estimated using Equation~\ref{eq:LF_errors}. The measured mass function is consistent with an unbroken power law within the estimated error bars.
%


\section{Summary and conclusion} \label{sec:conclusion}
Prior to the launch of JWST, all photometric studies of globular clusters were necessarily restricted to main sequence and post-main sequence stars, as even the deepest HST surveys lacked the sensitivity to unambiguously identify substellar members, \review{let alone measure the distribution of their parameters}. Now, the depth of NIRCam images affordable even with modest integration times allows not only to identify the brown dwarf cooling sequence in the CMD, but also to infer the extrinsic properties of the cluster such as its age and chemical composition from the observed color-magnitude trend \review{and luminosity function} beyond the end of the main sequence.

In this study, we identified three brown dwarf members of the globular cluster \ngc{} using new JWST NIRCam images in combination with archival HST ACS data from an earlier epoch for membership selection. While all three objects are distinctly visible in both NIRCam and ACS fields (Figure~\ref{fig:finders}), the cumulative integration time of the latter exceeds that of the former by multiple orders of magnitude. JWST is clearly well-suited for detection and characterization of brown dwarfs in globular clusters due to both its large aperture size, and sensitivity to infrared wavelengths near the peak of the spectral energy distribution of low-temperature metal-poor atmospheres.

The three newly discovered brown dwarfs, designated \textit{BD1756}, \textit{BD1628} and \textit{BD1388}, have been confirmed to be \textit{bona fide} substellar members of \ngc{} by extensive quality cuts in our photometric analysis, proper motion-based membership, direct visual inspection of the images and comparison with models. The estimated stellar parameters of the objects are listed in Table~\ref{tab:bd_params}. All identified brown dwarfs have masses below the metallicity-dependent hydrogen-burning limit, radii of $70-80\%$ of $R_\mathrm{jup}$, and probable spectral types of late L or early T. Other key findings of this work are summarized below:

\begin{enumerate}
    \item The lowest-mass main sequence stars in \ngc{} appear to be better approximated by more metal-rich isochrones than the upper main sequence and the turnoff point. This finding differs from our earlier analysis of the globular cluster \tuc{} \citep{roman_47Tuc}, and similar works in the literature \citep{JWST_low_MS_phot_1,JWST_low_MS_phot_2,JWST_low_MS_phot_3}, and may be indicative of unaccounted non-solar abundances of individual elements or imperfectly modelled low-temperature phenomena in cool atmospheres. A comparison of the theoretically predicted luminosity at the hydrogen-burning limit derived in this work and \citet{roman_47Tuc} with the candidate stellar/substellar gap in \citet{marino_47tuc_BD} further suggests that the effective oxygen abundance of brown dwarfs may be some $\sim\qty{0.5}{dex}$ lower than in higher-mass stars. This effect can be investigated further by introducing additional dimensions to the model grid, including the abundances of key chemical elements, and dust/cloud formation parameters. 
    %
    %
    A potentially related effect has been observed in stellar/substellar binary systems in the field \citep{BD_inconsistent_chemistry}.
    %
    \item The new grid of \texttt{SANDee} evolutionary models presented in this study allowed us to investigate the dependence of the hydrogen-burning limit on the chemical composition. We found that even the oldest globular clusters likely have significant numbers of objects just above the hydrogen-burning limit that are yet to reach the main sequence. These fast-evolving pre-main sequence stars alongside \textit{bona fide} brown dwarfs below the hydrogen-burning limit are particularly valuable in determining the ages of their parent clusters.
    \item The best-fit \texttt{SANDee} isochrone provides an excellent fit to both HST optical and JWST infrared CMDs of \ngc{} across the hydrogen-burning limit. \texttt{SANDee} is therefore the first grid of stellar models capable of accurately representing the stellar/substellar transition in globular clusters at a wide range of metallicities.
    \item Despite our analysis only involving three confirmed brown dwarf members, we were able to estimate the age of \ngc{} as $13.4\pm\qty{3.3}{Gyr}$ by fitting a model to the observed luminosity function of the cluster near the hydrogen-burning limit. This 
    result demonstrates a new dating method for globular clusters that is independent of other methods such as isochrone fitting around the turnoff point. The constraints on the inferred age are expected to improve as more brown dwarf members are characterized in future surveys. The estimated age was found to be largely independent of the adopted mass function, with the effect not exceeding $\qty{1}{Gyr}$ across the entire range of plausible mass function slopes.
    \item The present-day local mass function of \ngc{} within the considered field of view and range of stellar masses ($\qty{0.08}{\Msun}<M<\qty{0.18}{\Msun}$) was found consistent with a single-component power law without any statistically significant discontinuities across the hydrogen-burning limit. This result is expected, as both star formation and dynamical processes that shape the observed mass function should not depend on whether the cluster members undergo hydrogen fusion in their cores. This work is therefore a major improvement on the earlier analyses of the globular cluster mass functions near the hydrogen-burning limit (e.g., \citealt{rolly_M4_MF,ngc_6397_king}) that often found a prominent discontinuity below $\sim\qty{0.1}{\Msun}$, largely due to the systematic errors in the adopted mass-luminosity relationships.
    \item The top-heavy nature of the mass function inferred in this study serves as evidence of extensive dynamical evolution in \ngc{}. This finding is consistent with the short relaxation time of the cluster ($\sim\qty{0.4}{Gyr}$, \citealt{GC_distances}), and is further accentuated by the proximity of the chosen NIRCam field to the cluster core.
\end{enumerate}

The \texttt{SANDee} evolutionary models calculated in this study are made publicly available for future analyses of deep photometric surveys in globular clusters. We expect that far more precise estimates of globular cluster ages, mass functions and chemical abundances can be extracted from a larger sample of confirmed brown dwarf members. Since, at present, the photometric sensitivity of HST is the limiting factor in our analysis, a second epoch of observations with JWST within the same field is suggested as a natural follow-up study to this work. A visual inspection of the brown dwarf cooling sequence in the upper panel of Figure~\ref{fig:CMD}, as well as the estimated photometric completeness at the corresponding magnitudes from the lower panel of the figure suggest that the NIRCam field observed in this study may contain between $2$ and $3$ times more brown dwarfs than we were able to confirm, all at \texttt{F322W2} magnitudes brighter than $26$. This number of observable brown dwarfs is perhaps even larger at fainter magnitudes that could not be explored here due to the missing membership information, but are still within the sensitivity range of our JWST observations. It is therefore almost certain that another visit of the same field with JWST at a future epoch can vastly improve the statistical significance of our results, and unambiguously reveal the brown dwarf peak in the observed luminosity function.

    R.G. and A.J.B. acknowledge the funding support from Hubble Space Telescope (HST) Program GO-15096, provided by NASA through a grant from the Space Telescope Science Institute, which is operated by the Association of Universities for Research in Astronomy, Incorporated, under NASA contract NAS5-26555. 
    L.R.B. acknowledges 
    financial support by INAF under WFAP project, f.o.:1.05.23.05.05. 
    E.A. acknowledges the funding support of University of California Leadership Excellence through Advanced Degrees (UC LEADS). The computational demand of this study was met by the Advanced Cyberinfrastructure Coordination Ecosystem: Services \& Support (ACCESS), supported by the National Science Foundation. Some of the software used in this study was produced with assistance from ChatGPT 3.5. ChatGPT is a large language model developed and maintained by OpenAI.

\software{
\texttt{Astropy} \citep{astropy_1,astropy_2,astropy_3},  
\texttt{Matplotlib} \citep{matplotlib},
\texttt{NumPy} \citep{numpy}, 
\texttt{SciPy} \citep{scipy},
\texttt{MESA} \citep{MESA,MESA_2,MESA_3,MESA_4,MESA_5},
\texttt{BasicATLAS} \citep{mikaela},
\texttt{ChatGPT} \citep{ChatGPT}
}

\facilities{HST (ACS), JWST (NIRCam)}

\clearpage

\bibliography{references}{}
\bibliographystyle{aasjournal}

\end{document}